\documentclass[floatfix, twocolumn, prb]{revtex4}
\usepackage{hyperref}
\usepackage{color}
\usepackage{multirow}
\usepackage{epsfig}
\usepackage{graphicx}
\usepackage{amsmath}
\usepackage{amssymb}
\usepackage{bm}
\usepackage{dcolumn}
\usepackage{braket}
\usepackage{bbold}
\usepackage{ulem}
\begin{document}

%%%%%%%%%%%%%%%%%%%%%%%%%%%%%%%%%%%%%%%%%%%%%%%%%%%%%%%%%%%%%%%%%%%%%
%% The document title should be given as usual. Some journals require
%% a running title from the author: this should be supplied as an
%% optional argument to \title.
%%%%%%%%%%%%%%%%%%%%%%%%%%%%%%%%%%%%%%%%%%%%%%%%%%%%%%%%%%%%%%%%%%%%%

\title{Multifunctional Antiperovskites driven by Strong Magnetostructural Coupling}

\author{Harish K. Singh}
\author{Ilias Samathrakis}
\author{Nuno M. Fortunato}
\author{Jan Zemen}
\author{Chen Shen}
\author{Oliver Gutfleisch}
\author{Hongbin Zhang}
\affiliation{Institute of Materials Science, Technical University Darmstadt, Otto-Berndt-Strasse 3, 64287 Darmstadt, Germany}
\affiliation{Faculty of Electrical Engineering, Czech Technical University in Prague, Technicka 2, Prague 166 27, Czech Republic}
\affiliation{Department of Physics, Blackett Laboratory, Imperial College London, London SW7 2AZ, United Kingdom}
%\phone{+49 (0) 6151 16-23135}
%\email{harish@tmm.tu-darmstadt.de}
%\email{hzhang@tmm.tu-darmstadt.de}

%%%%%%%%%%%%%%%%%%%%%%%%%%%%%%%%%%%%%%%%%%%%%%%%%%%%%%%%%%%%%%%%%%%%%
%% Some journals require a list of abbreviations or keywords to be
%% supplied. These should be set up here, and will be printed after
%% the title and author information, if needed.
%%%%%%%%%%%%%%%%%%%%%%%%%%%%%%%%%%%%%%%%%%%%%%%%%%%%%%%%%%%%%%%%%%%%%

%%%%%%%%%%%%%%%%%%%%%%%%%%%%%%%%%%%%%%%%%%%%%%%%%%%%%%%%%%%%%%%%%%%%%
%% The manuscript does not need to include \maketitle, which is
%% executed automatically.
%%%%%%%%%%%%%%%%%%%%%%%%%%%%%%%%%%%%%%%%%%%%%%%%%%%%%%%%%%%%%%%%%%%%%
%\documentclass{article}

\begin{abstract}
 Based on density functional theory calculations, we elucidated the origin of multifunctional properties for cubic antiperovskites with noncollinear magnetic ground states, which can be attributed to strong isotropic and anisotropic magnetostructural coupling. 16 out of 54 stable magnetic antiperovskite M$_3$XZ (M = Cr, Mn, Fe, Co, and Ni; X = selected elements from Li to Bi except for noble gases and 4f rare-earth metals; and Z = C and N) are found to exhibit the $\Gamma_{4g}$/$\Gamma_{5g}$ ({\it i.e.}, characterized by irreducible representations) antiferromagnetic magnetic configurations driven by frustrated exchange coupling and strong magnetocrystalline anisotropy. Using the magnetic deformation as an effective proxy, the isotropic magnetostructural coupling is characterized, and it is observed that the paramagnetic state is critical to understand the experimentally observed negative thermal expansion and to predict the magnetocaloric performance. Moreover, the piezomagnetic and piezospintronic effects induced by biaxial strain are investigated. It is revealed that there is not a strong correlation between the induced magnetization and anomalous Hall conductivities by the imposed strain. Interestingly, the anomalous Hall/Nernst conductivities can be significantly tailored by the applied strain due to the fine-tuning of the Weyl points energies, leading to promising spintronic applications.
\end{abstract}

\maketitle
\section{Introduction}
Smart materials like multiferroic materials with enhanced coupling between different degrees of freedom ({\it e.g.}, mechanical, electronic, and magnetic) are promising for engineering devices for future applications such as sensors, transducers, memories, and spintronics.~\cite{fiebig2016evolution, trassin2015low, ma2011recent} Cubic antiperovskite (APV) compounds host the two most appealing aspects of multiferroics, {\it e.g.}, magnetoelectric coupling and piezomagnetic effect (PME).~\cite{spaldin2005renaissance, ma2011recent} In APV materials, the strong magnetoelectric coupling is achieved by combining piezoelectric and piezomagnetic heterostructure composites.~\cite{quintela2020epitaxial,Lukashev, shao2019electrically} 
A significant PME is reported for Mn-based nitrides like Mn$_3$SnN, making such compounds a suitable component for fabricating magnetoelectric composite.~\cite{Zemen2017PME, boldrin2018giant} The PME in APVs can be attributed to the strong magnetostructural coupling, which manifests itself also as giant negative thermal expansion (NTE)~\cite{lin2015giant, takenaka2014, NTE_Takenaka2005} and magnetocaloric/barocaloric effect.~\cite{matsunami2015, peng2013mn, shi2016baromagnetic, tohei2003negative, boldrin2018multisite, boldrin2018giant} From the materials perspective, many Mn-based APV carbides go through a first-order magnetic phase-transition and possess a large magnetocaloric effect.~\cite{peng2013mn, wang2009} For instance, Mn$_3$GaC exhibits a huge magnetic entropy change ($\Delta$S$_M$) of 15 J/kgK under an applied magnetic field of 2T.~\cite{tohei2003negative} The strong magnetostructural coupling in APVs is driven by the cubic-to-cubic first-order transition wherein a change in the crystal volume brings about a change in the frustrated magnetic states. Last but not least, APVs have been investigated recently due to the presence of a treasury of multifunctionality such as superconductivity,~\cite{he2001superconductivity} thermoelectric,~\cite{lin2014} magnetostriction.~\cite{shibayama2011giant} 
\\
Particularly, from the topological transport properties point of view, the existence of finite anomalous Hall conductivity (AHC) in noncollinear antiferromagnets has attracted noticeable attention due to possible applications in AFM spintronics for information storage and data processing.~\cite{chen2014anomalous, jungwirth2016antiferromagnetic, baltz2018antiferromagnetic, vzutic2004spintronics} The spin-dependent transport phenomena can provide spin-polarized charge current and large pure spin current, which could be achieved premised on two fundamental properties, {\it i.e.}, AHC and spin Hall conductivity (SHC). The kagome lattice turns out to be an elementary model to host giant AHC.~\cite{liu2018giant, xu2015intrinsic, wang2016anomalous} Recently, Mn-based APV nitrides have been proposed to exhibit large AHC in frustrated AFM  kagome-lattice.~\cite{Gurung, Samathrakis-Mn3GaN, boldrin2019anomalous, zhou2019spin} It is observed that Mn$_3$GaN exhibits vanishing and non-vanishing AHC for two different magnetic ordering $\Gamma_{5g}$ and $\Gamma_{4g}$, respectively.~\cite{Gurung, Samathrakis-Mn3GaN} In this regard, for magnetic materials with noncollinear AFM ground states, the non-vanishing AHC is only feasible with specific magnetic space group symmetry, {\it i.e.}, the AHC tensor depends on the magnetic group symmetry.~\cite{gallego2019automatic} For instance, Mn$_3$X (X= Ga, Ge, and Sn) and Mn$_3$Z (Z= Ir, Pt, and Rh) have been reported to have a different form of AHC tensor as a result of different magnetic ordering.~\cite{guo2017large, chen2014anomalous, kubler2014non, nayak2016large} A possible phase transition ($\Gamma_{5g}$$\leftrightarrow$$\Gamma_{4g}$) attained by strain or chemical modification could make these materials suitable for novel AFM spintronic applications. The spin-polarized current could also be generated by temperature gradient instead of the applied electric fields, resulting in anomalous Nernst conductivity (ANC), also termed as spin caloritronics.~\cite{bauer2012spin}$^,$~\cite{boona2014spin} 
A large ANC in noncollinear AFMs could be useful for establishing spin caloritronics devices that exhibit useful prospects in energy conversion and information processing. Noticeably, a large ANC of 1.80 AK$^{-1}$m$^{-1}$ has been reported for APV Mn$_3$NiN at 200 K,~\cite{zhou2020giant} which is slightly less than half of the highest reported ANC of 4.0 AK$^{-1}$m$^{-1}$ in Co$_2$MnGa.~\cite{sakai2018giant}$^,$~\cite{guin2019anomalous} Enhancing the AHC and ANC by applying strain could be crucial for realizing AFM spintronic  devices.    
\\
In this work, we carried out a systematic analysis of 54 cubic APV systems (Pm$\bar{3}$m) with chemical formula M$_3$XZ (see Fig. S1) to determine their magnetic ground states, their tunability via biaxial strain, and the resulting spintronic properties. Explicit calculations were performed to obtain the energies of four phases, {\it i.e.}, $\Gamma_{4g}$, $\Gamma_{5g}$, non-magnetic (NM), and ferromagnetic (FM)), where the magnetic anisotropy energy (MAE) defined as the energy difference between $\Gamma_{5g}$ and $\Gamma_{4g}$ is examined to understand the origin of the noncollinear magnetic states with the help of spin-orbit coupling energy. Moreover, a detailed analysis of the lattice constant variation with respect to the magnetic states reveals that the paramagnetic (PM) state is critical in the magnetic phase transition, enabling us to predict potential NTE and magnetocaloric materials. Last but not least, the PME was studied by introducing biaxial strains (compressive and tensile), which causes possible phase transitions between $\Gamma_{5g}$$\leftrightarrow$$\Gamma_{4g}$, and leads to a significant modification in the AHC and ANC, dubbed as a piezospintronic effect.~\cite{Samathrakis-Mn3GaN, liu2019antiferromagnetic} In-depth analysis on symmetry analysis and electronic structure suggests that the piezospintronic effect is originated from the existence of Weyl nodes whose position can be tailored by strain, resulting in promising applications for future spintronic devices.
\begin{figure}
	\begin{center}
		\includegraphics[width=0.5\textwidth]{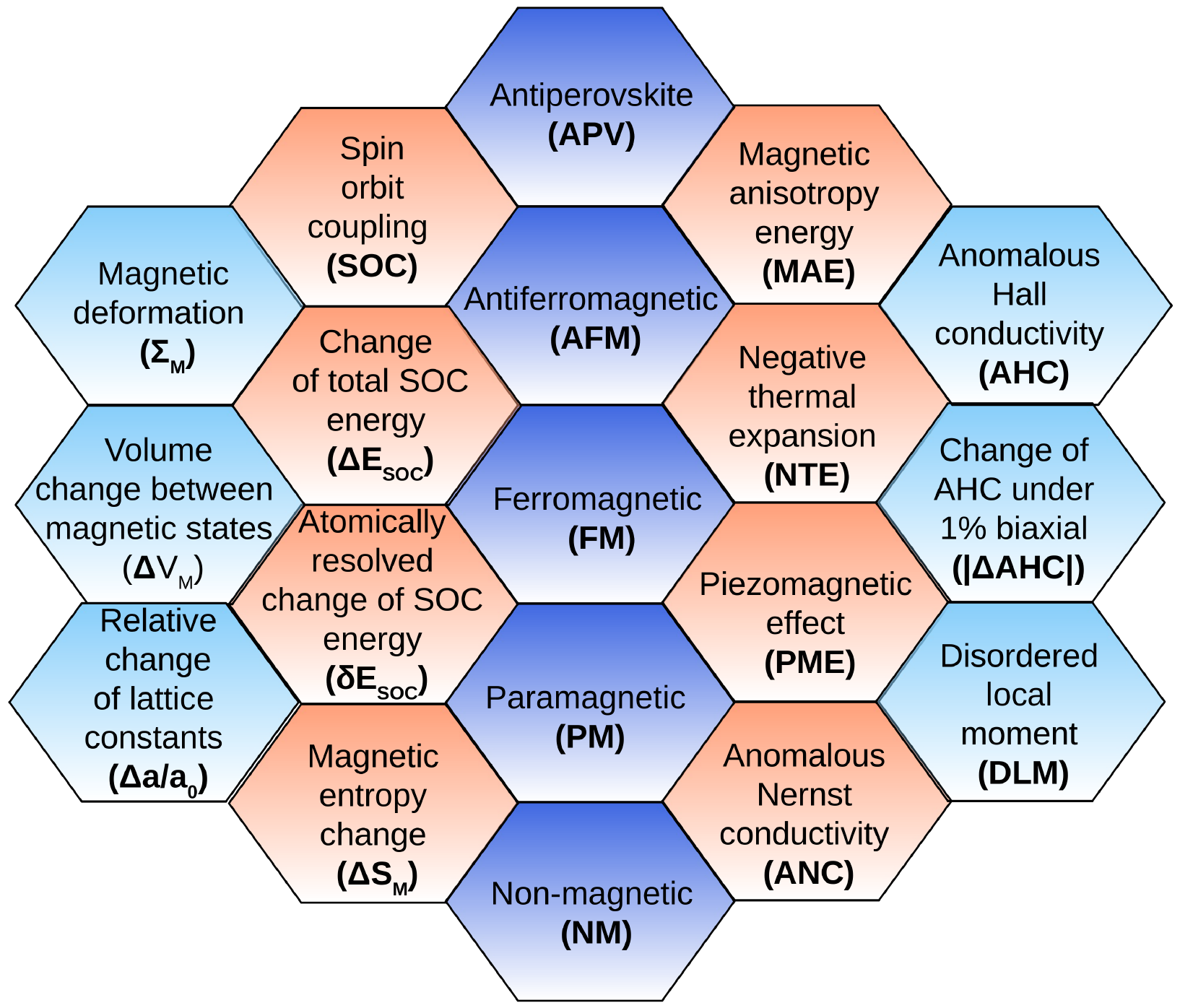}
		\caption{The list of abbreviations used in the manuscript.}
		\label{fig:magS}
	\end{center}
\end{figure}
\section{Computational Details}
Our density functional theory (DFT) calculations are performed using the projector augmented wave (PAW) method as effectuated in the VASP package.~\cite{kresse1996} The exchange-correlation functional is approximated using the generalized gradient approximation (GGA) as parameterized by Perdew-Burke-Ernzerhof (PBE).~\cite{perdew1996generalized} We used an energy cutoff of 500 eV for the plane-wave basis set, and a uniform k-points grid of 13$\times$13$\times$13 within the Monkhorst-pack scheme for the Brillouin zone integrations. The Methfessel-Paxton scheme is used to determine the partial occupancies of orbitals with a smearing width of 0.06 eV. The spin-orbit coupling (SOC) is considered in all the calculations. In order to verify the magnetic ground state of various compounds, the more accurate total energy calculations are performed using a higher energy cutoff of 600 eV and a dense k-mesh of 25$\times$25$\times$25.
\\
\begin{figure*}
	\begin{center}
		\includegraphics[width=1.0\textwidth]{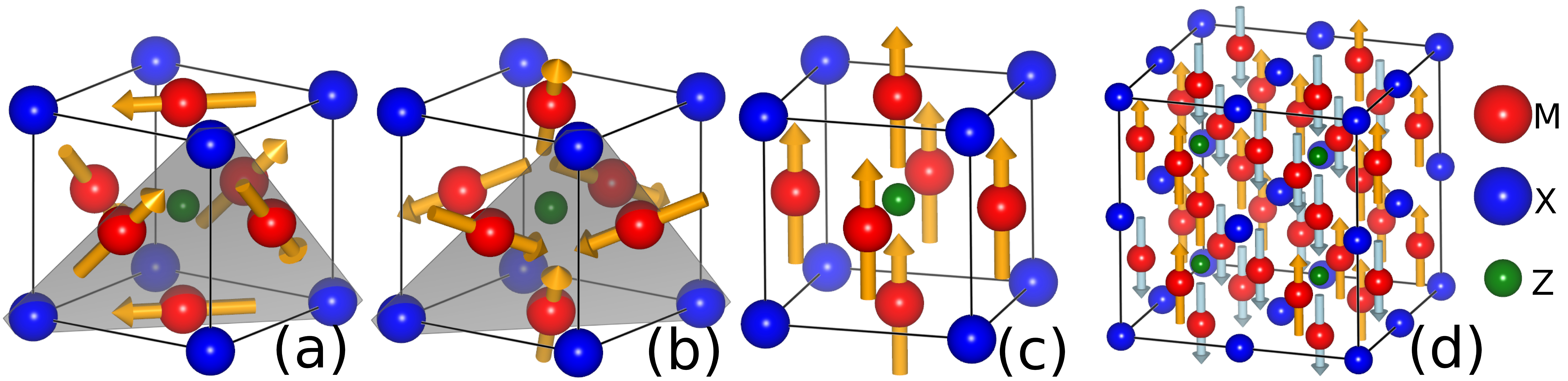}
		\caption{Possible magnetic structures of antiperovskites: (a) $\Gamma_{5g}$, (b) $\Gamma_{4g}$, (c) FM and, (d) PM. The spin chirality for $\Gamma_{4g}$ and $\Gamma_{5g}$ is assumed to be +1.}
		\label{fig:magS}
	\end{center}
\end{figure*}
The magnetic ground states are obtained by comparing the total energies of five configurations (see Table S1) $\Gamma_{4g}$, $\Gamma_{5g}$, FM, PM (see Fig.~\ref{fig:magS}, and NM states, where the lattice constants are fully optimized for each magnetic configuration (see Table S2). The PM state is modeled based on the disordered local moment (DLM) picture (see Fig.~\ref{fig:magS}(d)), where a special quasi-random structure modeled using a $2\times2\times2$ supercell by imposing zero total magnetization.~\cite{zunger1990special} That is, a supercell with the same number of up and down moments is considered with a pair correlation function same as A$_{50}$B$_{50}$ random alloys, generated using the alloy theoretic automated toolkit code.~\cite{avdw:atat} Moreover, to investigate the piezomagnetic and piezospintronic effects, biaxial in-plane strain is applied, which reduces the crystalline symmetry from cubic to tetragonal. 
The crystal structures are fully relaxed where the optimal lattice constants along c-direction are evaluated by the polynomial fitting of the energies from a series of calculations.
\\
The AHC is evaluated using the WannierTools code,~\cite{Wu2017}, where the required accurate tight-binding models are obtained by the maximally localized Wannier functions (MLWFs) using the Wannier90 code.~\cite{mostofi2008wannier90} 
The s, p, and d orbitals of M and X atoms and the  s and p orbitals for N atom are considered, resulting in 80 MLWFs in total for every noncollinear APVs. 
The AHC is computed by integrating the Berry curvature on a uniform 401$\times$401$\times$401 k-points mesh to guarantee good accuracy, which can be expressed as:~\cite{xiao2010berry}

\begin{equation}
\sigma_{xy} = - \frac{e^2}{\hbar} \int \frac{d\mathbf{k}}{\left( 2\pi \right)^3} \sum_{n} f \big[ \epsilon \left( \mathbf{k} \right) -\mu \big] \Omega_{n,xy} \left( \mathbf{k} \right) 
\end{equation} 
\begin{equation}
\Omega_{n,xy}\left( \mathbf{k} \right) = -2\text{Im} \sum_{m \neq n} \frac{{\braket{\psi_{\mathbf{k}n}|v_{x}|\psi_{\mathbf{k}m}}}{\braket{\psi_{\mathbf{k}m}|v_{y}|\psi_{\mathbf{k}n}}}}{\big[ \epsilon_m\left( \mathbf{k} \right) - \epsilon_n\left( \mathbf{k} \right) \big]^2} \end{equation}
%where, $\alpha$, $\beta$= x, y, z ($\alpha$ $\neq$ $\beta$),
where e is elementary charge, $\mu$ is the chemical potential, $\psi_{n/m}$ denotes the Bloch wave function with energy eigenvalue $\epsilon_{n/m}$,
$v_{x/y}$ is the velocity operator along Cartesian $x/y$ direction, and $f[\epsilon \left( \mathbf{k} \right) -\mu]$ is Fermi-Dirac distribution function. 
The ANC $\alpha_{xy}$ is evaluated based on the Mott relation, which yields:
\begin{equation}
\alpha_{xy} = - \frac{\pi^2k_B^2T}{3e} \frac {d\sigma_{xy}} {d\epsilon} \big |_{\epsilon=\mu} 
\end{equation}
In this work, we evaluated only the derivative of AHC at the Fermi level d$\sigma_{xy}$/d$\epsilon$, which provides a quantitative measurements of the ANC.
\section{Results and discussion}
\subsection{Validation and prediction of magnetic ground state}
As the magnetic transition metal atoms are located at the face centers of the cubic cell for magnetic APVs, it leads to a frustrated kagome-lattice in the (111)-plane. Correspondingly, noncollinear magnetic structures are expected when the interatomic exchange interaction is AFM for the nearest neighbors. As shown in Fig.~\ref{fig:magS},  $\Gamma_{4g}$, and $\Gamma_{5g}$ are the two most common magnetic configurations reported for APVs, resulting in 120$^\circ$ magnetic angle configurations within the (111)-plane between three magnetic moments. The $\Gamma_{4g}$ state can be obtained from $\Gamma_{5g}$ by simultaneously rotating the moments of three metal atoms within the (111)-plane by 90$^\circ$. The APVs with the noncollinear magnetic ground state are listed in Table.~\ref{tab:magS}, in comparison to the available experimentally measured results. Interestingly, all the APVs with noncollinear magnetic ground states are nitrides including Cr and Mn, while all carbides end up with the FM ground state except Mn$_3$SnC, which exhibits the $\Gamma$$_5$$_g$ state (see Table S1). It is noted that there are also other magnetic configurations such as ferrimagnetic and canted states for Mn$_3$XC (X = Ga, Sn, and Zn),~\cite{yu2003large, dias2015effect, antonov2007} which will be saved for future investigations.
\\
The magnetic ground states of noncollinear-APVs are in good agreement with the experimental measurements. For instance, both Mn$_3$GaN~\cite{matsunami2015} and Mn$_3$ZnN~\cite{fruchart, deng2015frustrated} have the $\Gamma_{5g}$ magnetic ground state, which are consistent with our DFT calculations. Interestingly, many Mn-based APVs exhibit a mixed $\Gamma_{4g}$ + $\Gamma_{5g}$ magnetic ordering, with a possible meta-magnetic transition to the other magnetic phases.~\cite{fruchart} For instance, Mn$_3$AgN is characterized by two distinct magnetic phase transitions. A mixed $\Gamma_{4g}$ + $\Gamma_{5g}$ phase exists below 55 K, whereas pure $\Gamma_{5g}$ state persists at intermediate temperature range (55 K$<$ T $<$290 K), and lastly, it undergoes the magnetic transition to PM state at 290 K.~\cite{fruchart} Likewise, Mn$_3$NiN displays such mixed magnetic phases.~\cite{fruchart} Mn$_3$SnN shows four different magnetic and crystallographic phases.~\cite{fruchartMn3SnN}
\begin{table*}
	\caption{Compilation of the APVs (M$_3$XZ) with noncollinear magnetic ground state. All the units are in meV. For each compound, the calculated magnetic ground state (``magGS'') is listed and the change in the atomically resolved SOC energy ($\delta$E$_\text{SOC}$) between $\Gamma_{5g}$ and $\Gamma_{4g}$ state are enumerated for M and X atoms. The calculated MAE and $\Delta$E$_\text{SOC}$ are summarized as given by equation ~\ref{eq:MAE} and ~\ref{eq:SOC}, respectively. The last column lists the available experimental magnetic ground state.}
	\label{tab:magS}
	\begin{tabular}{|c|c|c|c|c|c|c|}
		\hline\centering
		M$_3$XZ&magGS&\textsuperscript{M}($\delta$E$_\text{SOC}$)&\textsuperscript{X}($\delta$E$_\text{SOC}$)&$\Delta$E$_\text{SOC}$&MAE&Refs.\\[4pt] 
		\hline\centering 
		Cr$_3$IrN & $\Gamma_{4g}$&-0.367&4.532&4.164&2.187 & \\
		Cr$_3$PtN & $\Gamma_{4g}$&-0.936&5.639&4.703&2.868& \\
		Cr$_3$SnN & $\Gamma_{4g}$&-0.568&1.158&0.589&0.471 & \\
		Mn$_3$AgN & $\Gamma_{4g}$&0.3987&-0.2486&0.151&0.063 &$\Gamma$$_{4g}$+$\Gamma$$_{5g}$~\cite{fruchart} \\
		Mn$_3$AuN & $\Gamma_{5g}$&1.205&-4.353&-3.148&-1.671 & \\
		Mn$_3$CoN & $\Gamma_{4g}$&0.005&1.716&1.721&1.218 & AFM~\cite{takenaka2014}\\
		Mn$_3$GaN & $\Gamma_{5g}$&-1.111&1.10&-1.110&-0.291 & $\Gamma_{5g}$~\cite{fruchart, Bertaut1968} \\
		Mn$_3$HgN & $\Gamma_{5g}$&-0.746&-3.691&-4.438&-1.551 & \\
		Mn$_3$InN & $\Gamma_{5g}$&-0.734&-0.224&-0.958&-1.849 & \\
		Mn$_3$IrN & $\Gamma_{4g}$&0.599&19.705&20.305&10.548 & \\
		Mn$_3$NiN & $\Gamma_{4g}$&1.216&-0.87&0.346&0.144 &$\Gamma$$_{4g}$+$\Gamma$$_{5g}$~\cite{fruchart}\\
		Mn$_3$PdN & $\Gamma_{4g}$&1.292&-1.115&0.176&0.357 & AFM~\cite{takenaka2014}\\
		Mn$_3$PtN &$\Gamma_{5g}$&3.872&-13.877&-10.005&-4.969 & \\
		Mn$_3$RhN &$\Gamma_{4g}$&-0.021&3.482&3.460&1.702 & \\
		Mn$_3$SnN & $\Gamma_{4g}$&0.099&0.479&0.578&0.190 &$\Gamma$$_{4g}$+$\Gamma$$_{5g}$~\cite{fruchart}\\
		Mn$_3$ZnN & $\Gamma_{5g}$&-0.756&-0.105&-0.861&-1.452 & $\Gamma_{5g}$~\cite{fruchart}$^,$~\cite{deng2015frustrated} \\
		\hline
	\end{tabular}
\end{table*}
\\
The mixed magnetic ordering can be attributed to the MAE between the $\Gamma_{5g}$ and $\Gamma_{4g}$ states, which can be expressed as
\begin{equation}
\label{eq:MAE}
\text{MAE}= \text{E}_{\Gamma_{5g}}-\text{E}_{\Gamma_{4g}},
\end{equation}
where $\text{E}$ indicates the total energy of the corresponding magnetic configuration. For Mn$_3$AgN, Mn$_3$NiN, and Mn$_3$SnN with experimentally observed mixed magnetic states, the corresponding MAE is 0.063 meV, 0.144 meV, and 0.190 meV, respectively. Nevertheless, it is still unclear why Mn$_3$AgN is stabilized in the $\Gamma_{5g}$ state between 55K and 290K. We suspect that those compounds with MAE greater than 0.2 meV between $\Gamma_{4g}$ and $\Gamma_{5g}$ should have pure noncollinear magnetic states. 
In our study, the Mn-based APVs Mn$_3$XN with X = Ag, Co, Ir, Ni, Pd, Rh, and Sn exhibit $\Gamma_{4g}$ while those with X = Au, Ga, Hg, In, Pt, and Zn display the $\Gamma_{5g}$ as the magnetic ground state (see Table~\ref{tab:magS} and S1). This is consistent with the recent DFT calculations for Mn$_3$XN (X = Ni, Zn, Ga, Sn, and Pt)~\cite{huyen2019topology}, except for Mn$_3$XN (X= In, Pd, and Ir).  
In Ref. 51, Mn$_3$InN exhibit $\Gamma_{4g}$ state while Mn$_3$XN (X= Ir and Pd) display $\Gamma_{5g}$ state. The MAE for Mn$_3$InN calculated by Huyen {\it et al.} is 74.6 meV, which is much larger than our MAE (-1.84 meV). As detailed below, a large MAE could be expected for the compound with strong SOC; thus, it is questionable for Mn$_3$InN with such a large MAE of 74.6 meV.~\cite{huyen2019topology}  To clarify the discrepancy, we performed calculations of the MAE using the Quantum Espresso (QE) code~\cite{giannozzi2009quantum} with the same parameters consider in Huyen {\it et al.} study. The QE and VASP calculations provide the same magnetic ground state and MAE for Mn$_3$XN (X= Ir and Pt) systems, which is in contradiction to the Huyen study performed using QE.~\cite{huyen2019topology} 
\\
Lastly, based on our high-throughput calculations, three unreported Cr-based APVs are stable, our calculations reveal that Cr$_3$XN (Cr$_3$IrN, Cr$_3$SnN, and Cr$_3$PtN) have $\Gamma_{4g}$ as the lowest energy magnetic configuration. While among APV nitrides, Mn$_3$AuN and Mn$_3$HgN are the other two new noncollinear systems with $\Gamma_{5g}$ magnetic ground state. To the best of our knowledge, there is no experimental neutron diffraction measurement available til now for such APVs, making them interesting for future studies. To shed more light on the origin of MAE, we performed detailed calculations of the SOC energy in order to obtain atomically resolved contributions to the MAE. Like MAE, the change in the total SOC energy ($\Delta$E$_\text{SOC}$) of M$_3$XN can be defined as the difference between the SOC energy (E$_\text{SOC}$) of the $\Gamma_{5g}$ and $\Gamma_{4g}$ magnetic states.~\cite{antropov2014constituents}
\begin{equation}
\label{eq:SOC}
\Delta\text{E$_\text{SOC}$}= \text{E}_\text{SOC} ({\Gamma_{5g}})-\text{E}_\text{SOC}({\Gamma_{4g}}),
\end{equation}
In the same way, the atomically resolved change of SOC energy ($\delta$E$_\text{SOC}$) can also be defined for each atomic species, as shown in Table~\ref{tab:magS}. The positive (negative) values of MAE and $\Delta$E$_\text{SOC}$ indicate the $\Gamma_{4g}$ ($\Gamma_{5g}$) magnetic ground state. On the one hand, it is found that the sign of MAE is the same as that of $\Delta$E$_\text{SOC}$ (see Table~\ref{tab:magS}). That is, both the $\Delta$E$_\text{SOC}$ and MAE are equally valid to characterize the magnetic ground states, with a linear scaling behavior observed (see Fig. S2). For example, Cr$_3$PtN has $\Gamma_{4g}$ magnetic ground state with an MAE of 2.868 meV and $\Delta$E$_\text{SOC}$ of 4.70 meV. The atomic resolved $\delta$E$_\text{SOC}$ of Pt and Cr are 5.639 meV and -0.938 meV. Evidently, the $\delta$E$_\text{SOC}$ of Pt is larger than Cr $\delta$E$_\text{SOC}$ originating in the strong SOC of Pt. As a result, the $\delta$E$_\text{SOC}$ of Pt is a determining factor for the MAE and $\Delta$E$_\text{SOC}$ sign. On the other hand, for most APVs, the contribution of the X elements to MAE is more significant than that of the magnetic elements M, as indicated by the atomic resolved $\delta$E$_\text{SOC}$ (see Table~\ref{tab:magS}). Taking the Mn$_3$XN with X = Co, Rh, and Ir as examples, the MAE are 1.22 meV, 1.70 meV, and 10.55 meV, corresponding to the dominant $\delta$E$_\text{SOC}$ of X as 1.72 meV, 3.46 meV, and 20.31 meV, respectively. It is noted that $\delta E_\text{SOC}$ of Mn is smaller than 3\% of $\Delta E_\text{SOC}$ for such compounds. This can be attributed to the enhanced strength of atomic SOC of the X elements, {\it e.g.}, the strength of atomic SOC for Co, Rh, and Ir atoms is about 0.065 eV, 0.152 eV, and 0.452 eV, respectively.~\cite{buschow2003handbook} Therefore, the enhanced atomic SOC strength of the X elements is favorable for the strong MAE of the APV compounds.
\subsection{NTE and barocaloric}
\begin{figure}[h]
	\begin{center}
		\includegraphics[width=0.5\textwidth]{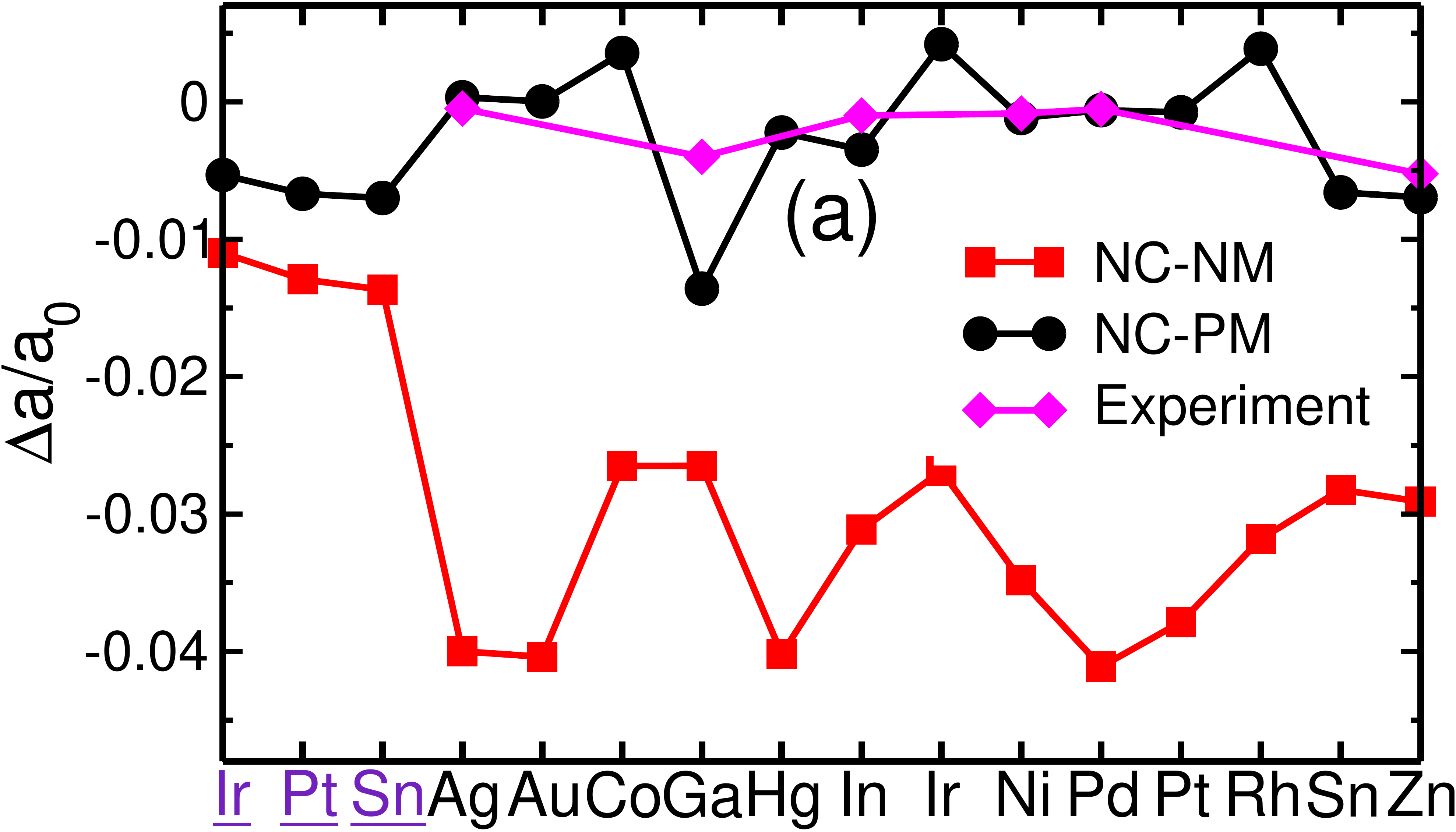}
		\includegraphics[width=0.49\textwidth]{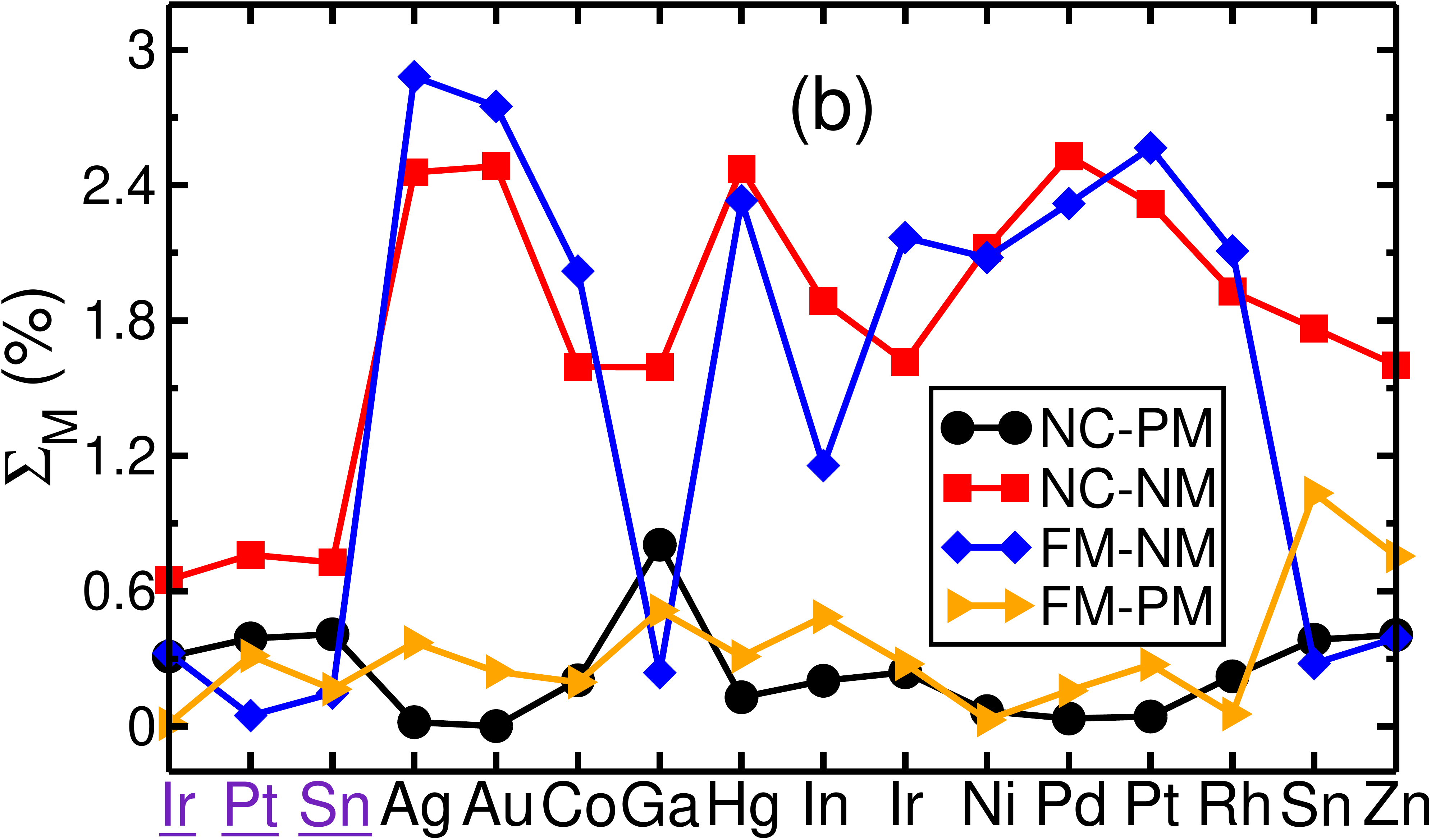}
		\caption{(a) The relative change in the lattice constant ($\Delta$a/a$_0$) and (b) the magnetic deformation $\sum_{M}$ for the Cr$_3$\underline{X}N and Mn$_3$XN antiperovskites (\underline{X} and X are the elements present on the x-axis in violet and black, respectively). The lattice constants of noncollinear, FM, PM, and NM states are used to determine $\Delta$a/a$_0$ and $\sum_{M}$.}
		\label{fig:lattice-caloric}
	\end{center}
\end{figure}

Turning now to the magnetostructural coupling, which can be best represented for APVs by the negative thermal expansion (NTE), magnetocaloric, and PME. 
NTE materials exhibit contraction in the lattice parameters with respect to temperature in contrast to most materials.\cite{Shio2015} 
To study NTE of APVs, we evaluated the relative change in the lattice constants $\Delta$a/a$_0$ on the transition from the noncollinear magnetic ground state to PM state (where a$_0$ is the lattice constant of $\Gamma_{4g}$ or $\Gamma_{5g}$ noncollinear state and $\Delta$a is lattice constant difference between PM and $\Gamma_{4g}$ or $\Gamma_{5g}$ state) and compared with the $\Delta$a/a$_0$ values obtained from the experimental measurement at their transition temperature (see Fig.~\ref{fig:lattice-caloric}(a)).~\cite{takenaka2014, NTE_Takenaka2005, Shio2015} The average of the evaluated $\Delta$a/a$_0$ is in good agreement with the experimental observation except for Mn$_3$GaN (see Fig.~\ref{fig:lattice-caloric}(a)). Together with the mismatch for the other cases, we suspect it may be due to the finite temperature effect, as our DFT calculations are done at zero Kelvin.
\\
Importantly, it is found that the PM state is critical and cannot be approximated as the NM state, {\it e.g.}, the $\Delta$a/a$_0$ between the noncollinear and NM states is about five times as large as the experimental value (see Fig.~\ref{fig:lattice-caloric}(a)). An ameliorated estimation could be achieved by defining the PM phase as a collinear AFM arrangement generated by using the disordered local moment (DLM) theory.~\cite{abrikosov2016recent} The $\Delta$a/a$_0$ of Cr$_3$IrN, Cr$_3$PtN, and Cr$_3$SnN on the transition from $\Gamma_{4g}$ magnetic ground state to PM  are -0.0053, -0.0067, and -0.0070, respectively. That is, Cr-based APVs undergo the lattice contraction with $\Delta$a/a$_0$  comparable to the reported NTE in Mn-based APV materials such as Mn$_3$GaN and Mn$_3$ZnN (see Fig.~\ref{fig:lattice-caloric}(a)). Furthermore, the equilibrium lattice constants decrease and increase for the Mn$_3$XN (X=Ga, Hg, In, Ni, Pd, Pt Sn, and Zn) and Mn$_3$XN APVs (X = Au, Co, Ir, and Rh) on the transition from a noncollinear magnetic ground state to PM state, respectively (see Fig.~\ref{fig:lattice-caloric}(a)).
\\
Recently, Bocarsly {\it et al.} proposed a magnetic deformation proxy $\sum_{M}$ to predict the magnetocaloric materials as the theoretical evaluation of ($\Delta$S$_M$) is very challenging.~\cite{Bocarsly2017} This proxy measures the volume deformation of the NM and magnetic structure, which correlates well with the experimentally measured $\Delta$S$_M$ values. In their study, they screened 167 ferromagnetic materials, including seven APVs. For the APVs, it is noted that $\sum_{M}$ scales linearly with the percentage volume difference $\Delta$V$_M$ between the FM and NM unit cell (see Fig. S3). Inspired by their work, we evaluated $\sum_{M}$ for the APVs with FM and noncollinear magnetic ground state in order to estimate the potential magnetocaloric effect. For FM APVs, it is found that the $\sum_{M}$ is most significant for the Fe-based APVs within the 1.2-1.7\% range (see Table S5). For example, the $\sum_{M}$ of Fe$_3$RhN, Fe$_3$PtN, and Fe$_3$NiN are 1.67\%, 1.59\%, and 1.54\%, respectively. Whereas Co-, Mn-, and Ni-based APVs have $\sum_{M}$ $<$ 1.0\% except for Mn$_3$ZnC (1.33\%) (see Table S5). Thus, we expect a significant magnetocaloric effect in Fe$_3$RhN, Fe$_3$PtN, and Fe$_3$NiN, based on the argument that the $\sum_{M}$ approximate to 1.5\% could exhibit a considerable magnetocaloric effect.~\cite{Bocarsly2017} 
\\
As to the noncollinear APVs, we evaluated the $\sum_{M}$ by considering four $\Delta$V$_M$ combinations, namely, the noncollinear magnetic ground state with PM and NM states (NC-PM and NC-NM), the FM with PM and NM states (FM-PM and FM-NM), see Fig.~\ref{fig:lattice-caloric}(b)). It is observed that the $\sum_{M}$ value is small by using the $\Delta$V$_M$ of the NC-PM and FM-PM states, in comparison to the other two combinations (see Fig.~\ref{fig:lattice-caloric}(b)). Nevertheless, a small value of $\sum_{M}$ does not mean a low $\Delta$S$_M$.~\cite{Bocarsly2017} For instance, the  experimentally measured $\Delta$S$_M$ of Mn$_3$GaN is very significant (22.3 Jkg$^{-1}$K$^{-1}$), and it also exhibits giant barocaloric effect.~\cite{matsunami2015} Overall, the  $\sum_{M}$ of Mn$_3$GaN should be largest among all the APVs. The calculated $\sum_{M}$ of Mn$_3$GaN is largest 0.805\% for the $\Gamma_{5g}$ and PM state combination, which verifies the experimental observation. Whereas, the $\sum_{M}$ of Mn$_3$GaN for the other three combinations is not the largest among the APVs (see Fig.~\ref{fig:lattice-caloric}(b)). Therefore, it can be concluded for the noncollinear systems, that the approach of Bocarsly {\it et al.} proxy does not provide reasonable $\sum_{M}$ obtained from the $\Delta$V$_M$ combination of NC-NM, FM-NM, and FM-PM states. The implicit  correlation of $\Delta$S$_M$ with $\sum_{M}$ indicates the strong magnetostructural coupling in APVs with significant $\Delta$V$_M$ (NC-PM). This conduces to the prediction of possible NTE and magnetocaloric materials with sizable $\sum_{M}$ by virtue of significant $\Delta$V$_M$ (see Table S6). For example, the Cr$_3$XN APVs (X= Ir, Pt, and Sn) with appreciable $\sum_{M}$ could exhibit potential applications as NTE and magnetocaloric materials.
\subsection{Piezomagnetism}
PME provides another effective characterization of the magnetostructural coupling,~\cite{Lukashev} which manifests itself as the response of magnetization to strain. The origin of PME is explained based on symmetry together with an illustration of the magnetic spin directions (see Fig. S4 and the Piezomagnetism section in the supplementary information). For APVs with noncollinear magnetic ground states, the total bulk magnetization is vanishing, but a net magnetization can be induced under finite compressive/tensile strain, as reported for Mn-based APVs.~\cite{Lukashev, Zemen2017PME} Our calculations confirm their results on the Mn-based APVs except for the Mn$_3$CoN and Mn$_3$RhN (see Fig. S5), where the resulting net magnetization from our (Zemen {\it et al.})~\cite{Zemen2017PME} calculations are 0.646 (0.305) and 0.214 (-0.143)  $\mu$$_B$/f.u., respectively. Interestingly, for Mn$_3$CoN, a net magnetic moment of 0.362 $\mu$$_B$/f.u. is induced at Co atoms under 1\% tensile strain, whereas the local magnetic moment of Co is zero in the cubic unstrained state. The magnetic moment direction of Co is aligned in (111) plane similar to Mn atom direction in the $\Gamma_{4g}$ state. As a result, Mn$_3$CoN could be an interesting material for elastocaloric and magneto-elastic applications. Also, newly predicted Cr$_3$PtN exhibits significant PME as large as 0.21 $\mu$$_B$/f.u. at 1\% compressive strain. Surprisingly, the PME effect is asymmetric with respect to the applied compressive and tensile strains. Most APVs exhibit more significant PME with tensile strain except Mn$_3$SnN and Cr$_3$PtN (see Fig. S6 and S7), {\it e.g.}, the net magnetization for Mn$_3$SnN at 1\% compressive strain is as large as 0.73 $\mu$$_B$/f.u. in comparison to 0.47 $\mu$$_B$/f.u. at 1.0\% tensile strain.
\\
The magnitude of the magnetoelectric effect is minimal in the intrinsic bulk AFM materials, such as Cr$_2$O$_3$.~\cite{Date1961}. The two-phase heterostructure materials consolidate the magnetoelectric effect. A recent study corroborates APV as a potential material for magnetoelectric composite.~\cite{Lukashev, Ryu2002} The piezoelectric perovskite could be one suitable substrate for such composite heterostructure as they also have comparable lattice parameters with APVs.~\cite{quintela2020epitaxial, Tashiro2013} The heterostructure amalgamate the piezoelectric and piezomagnetic properties, which are coupled by an interfacial strain. Based on  our PME analysis, we propose Cr- and Mn-based noncollinear systems with significant PME as potentials candidates for magnetoelectric composite. 
\\
Interestingly, the biaxial strain can induce phase transition between different magnetic configurations, {\it i.e.}, $\Gamma_{4g}$ and $\Gamma_{5g}$), which can further lead to a significant change in the transport properties as discussed below. It is found that a few APV materials undergo a magnetic phase transition due to the imposed biaxial strain (see Table S4). For instance, cubic Cr$_3$SnN has $\Gamma_{4g}$ as the magnetic ground state, which preserves under compressive strain. However, a transition into a distorted $\Gamma_{5g}$ state is obtained by applying a tensile strain of 0.5\% and 1.0\%. Similarly, there is a phase transition for Mn$_3$AuN from $\Gamma_{5g}$ to $\Gamma_{4g}$ driven by compressive strain, whereas the $\Gamma_{5g}$ configuration is retained under the tensile strain (see Table S4). Such transitions can be attributed to significant changes in the MAE with respect to the biaxial strain (see Table S4), but there is no consistent MAE trend as the absolute value of MAE is mostly determined by the atomic SOC strengths.
\subsection{Anomalous Hall Conductivity (Unstrained)}
It is well known that FM compounds commonly exhibit AHC due to the presence of a net magnetization,~\cite{nagaosa2010anomalous} where the broken time-reversal symmetry ($\mathcal{T}$) and SOC are two essential prerequisites. Noncollinear AFMs also display non-zero AHC, as observed in Mn$_3$X (X = Ir, Ge, and Sn).~\cite{guo2017large, chen2014anomalous} As a linear response property, the occurrence of AHC can be elucidated based on the symmetry analysis. For cubic APVs, the magnetic space group of $\Gamma_{5g}$ and $\Gamma_{4g}$ are $R\bar{3}m$ (166.97) and $R\bar{3}m'$ (166.101), respectively. For $\Gamma_{5g}$, the local magnetic moment spin directions of M atoms are invariant under the mirror symmetry transformation $M$$_{(0\bar11)}$, $M$$_{(10\bar1)}$, and $M$$_{(\bar110)}$, leading to vanishing AHC (see Table S8).~\cite{Samathrakis-Mn3GaN, Gurung} Whereas in the $\Gamma_{4g}$ state, the above-mentioned mirror symmetries are broken, but the product of mirror ($M$) and time-reversal symmetry $\mathcal{T}$ retain the $\Gamma_{4g}$ configuration. This results in finite AHCs with all three off-diagonal components non-zero but of the same amplitude (see Table S8).~\cite{Gurung} 
\\
Explicit calculations confirm the above symmetry arguments. For instance, the AHC is zero for Mn$_3$XN (X = Au, Hg, In, Pt, and Zn) with the $\Gamma_{5g}$ ground state (see Fig. S8(b) and S9). Whereas, the significant AHC is observed for the APVs with the $\Gamma_{4g}$ ground state, such as the AHC of Cr$_3$IrN and Cr$_3$PtN are 414.6 and 278.6 S/cm, respectively. The AHC of APVs is summarized in the supplementary information (see Fig. S8, S9, and S10). In comparison to the previous studies on an individual or a few compounds,~\cite{zhou2019spin, Gurung} our results are in good agreement (see Table S9). For instance, the AHC of Mn$_3$SnN from our and calculations by Gurung {\it et al.} are 106.5 and 133 S/cm~\cite{Gurung}, respectively. Whereas the value (-73.9 S/cm) obtained by Huyen {\it et al.}~\cite{huyen2019topology} is of opposite sign in addition to the difference in the absolute values. In this regard, for Mn$_3$NiN, the sign of experimentally observed AHC is confirmed by our DFT calculations,~\cite{boldrin2019anomalous} consistent with that obtained by Zhou {\it et al.}~\cite{zhou2019spin, zhou2020giant} This might be due to opposite magnetization direction while maintaining the noncollinear configurations or chirality of noncollinear spin configurations specified in the calculations, subject to further investigations. Moreover, it is observed that the absolute values of AHC for the same compound in the same magnetic ground state are scattered as well (see Table S9). We suspect that this is due to the different lattice constants and numerical parameters used in different calculations. 
For instance, Zhou {\it et al.} performed calculations using the experimental lattice constants on Mn$_3$NiN and other APV compounds, which are 
on average about 1.0-1.3\% smaller than the fully relaxed lattice constants in this work.
\begin{figure*}
	\begin{center}
		\includegraphics[width=1.0\textwidth]{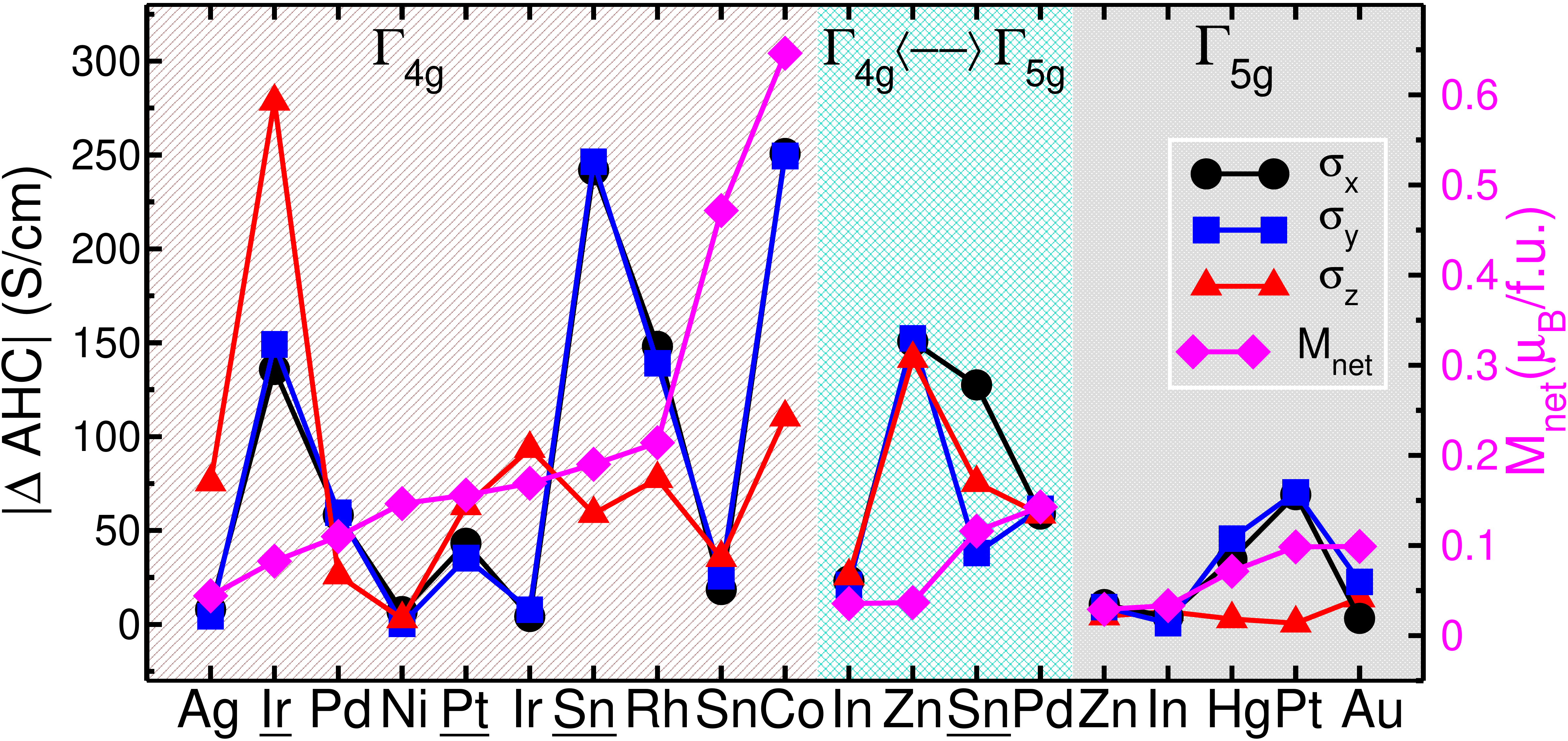}
		\caption{The piezospintronic effect is illustrated by the change in AHC ( $|$$\Delta$AHC$|$) at 1\% tensile strain and cubic phase. The $|$$\Delta$AHC$|$ and net magnetization is shown for the Cr$_3$\underline{X}N and Mn$_3$XN antiperovskites (\underline{X} and X are the elements present on the x-axis, respectively). The left, middle, and right panels correspond to the $\Gamma_{4g}$, phase transition between $\Gamma_{4g}$ and $\Gamma_{5g}$, and $\Gamma_{5g}$ state, respectively.}
		\label{fig:Piezospintronic}.  
	\end{center}
\end{figure*} 
\subsection{Strain induced AHC (Piezospintronics)}
As discussed above, that all cubic APVs with noncollinear magnetic ground states display significant PMEs where a net magnetization can be induced by applying biaxial strain; an interesting question is whether the AHC can be tailored as well. From the symmetry point of view, for the $\Gamma_{5g}$ state, the magnetic space group changes from $R\bar{3}m$ (166.97) to $C2/m$ (12.58) with biaxial strain, resulting in finite $\sigma_x$ and $\sigma_y$ with the same amplitude but opposite sign, while the $\sigma_z$ component remains zero.~\cite{Samathrakis-Mn3GaN} This is confirmed by our calculations on those compounds with the $\Gamma_{5g}$ ground state. For instance, the AHC induced by 1\% biaxial strain is as large as -71.2 S/cm for Mn$_3$PtN (see Fig. S8(b)), comparable to that in Mn$_3$HgN (see Fig. S9). The Mn$_3$InN and Mn$_3$ZnN have quite low AHC due to weak spin-orbit coupling strength (see Fig. S9).
\\
Similarly, the AHC of APVs in the $\Gamma_{4g}$ state can also be tailored by the biaxial strain. In this case, the magnetic space group is reduced from $R\bar{3}m'$ (166.101) to $C2'/m'$ (12.62), the resulting $\sigma_x$ and $\sigma_y$ components are the same while the $\sigma_z$ component possesses a distinct value (see Fig. S8(a), and Table S8). Explicit evaluations of the AHC for APVs in distorted $\Gamma_{4g}$ states verify the symmetry arguments, where 
the $\sigma_x$ and $\sigma_y$ components behave the same with respect to strain (see Fig. S8(a)). Interestingly, it is observed that more significant changes occur in AHC for the $\Gamma_{4g}$ compounds than the $\Gamma_{5g}$ cases. For instance, the $\sigma_z$ component of AHC in Cr$_3$IrN attains the largest AHC of 693.1 S/cm at 1\% tensile strain, with an increase of  278 S/cm compared to the value in the cubic geometry (see Fig. S8(a)). Moreover, the moderate strain can even lead to a sign change of the AHC, as observed on the $\sigma_z$ component of Mn$_3$AgN and Mn$_3$CoN with 0.5\% compressive and 1\% tensile strain, respectively (see Fig. S10).
\\
One interesting question is whether the piezospintronic effect ({\it i.e.}, the variation of AHC induced by biaxial strain) correlates with the PME, as summarized in Fig.~\ref{fig:Piezospintronic} for the variation of such quantities comparing the cases with 1\% tensile strain and cubic geometries. Obviously, $|$$\Delta$AHC$|$ cannot be directly inferred from the induced net magnetization, particularly exemplified by the APVs in the $\Gamma_{4g}$ magnetic ground state. For example, the $|$$\Delta$AHC$|$ and net magnetization of Cr$_3$IrN are 278.5 S/cm ($\sigma_z$ component) and 0.082 $\mu$$_B$/f.u., respectively, while for Mn$_3$SnN, the $|$$\Delta$AHC$|$ and net magnetization are 35.5 S/cm ($\sigma_z$ component) and 0.47 $\mu$$_B$/f.u., respectively. Moreover, for Mn$_3$CoN, the net magnetization and $|$$\Delta$AHC$|$ of $\sigma_x$ are 250.7 S/cm and 0.646 $\mu$$_B$/f.u. It signifies that the $|$$\Delta$AHC$|$ could be large for the small net magnetization and vice-versa (see Fig.~\ref{fig:Piezospintronic}). Interestingly, the $|$$\Delta$AHC$|$ and net magnetization are smaller for those compounds in the $\Gamma_{5g}$ magnetic state, in comparison to those of such materials with the $\Gamma_{4g}$ magnetic state, {\it e.g.}, the $|$$\Delta$AHC$|$ and net magnetization of Mn$_3$ZnN are 9.17 S/cm and 0.029 $\mu$$_B$/f.u., respectively.
\\
\begin{figure*}
	\begin{center}		
		\includegraphics[width=1.0\textwidth]{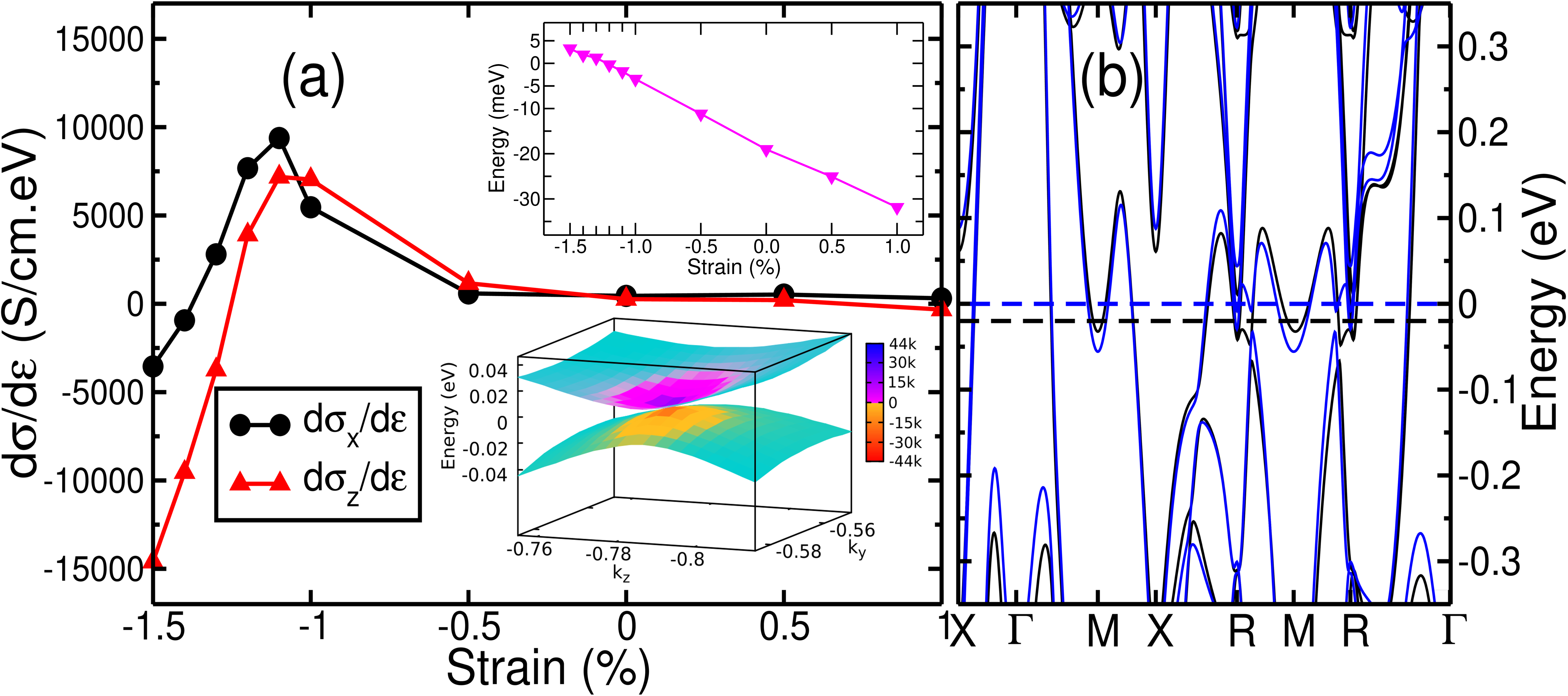}
		\caption{(a) The calculated ANC of Mn$_3$PdN in the $\Gamma_{4g}$ magnetic state. The Weyl points  (0.43, -0.35, -0.49 and equivalent coordinates) shift across the Fermi level with strain (inset above) and the Berry curvature for the 1.4\% compressive biaxial strain (inset below). (b) The band structure of Mn$_3$PdN for unstrained (black) and 1.4\% compressive biaxial strained (blue). The presence of Weyl point near the Fermi energy is indicated by black (cubic) and blue dashed line (-1.4\% biaxial strain).}
		\label{fig:ANE}
	\end{center}
\end{figure*}
Particularly, the APVs undergoing a phase transition between $\Gamma_{4g}$ and $\Gamma_{5g}$ states driven by the epitaxial strain exhibit strong piezospintronic effect while the magnitude of the PME is marginal. For instance, Mn$_3$ZnN has a $\Gamma_{5g}$ magnetic ground state in the cubic phase with vanishing AHC. A phase transition into the $\Gamma_{4g}$ state by 1\% tensile strain results in a considerable $|$$\Delta$AHC$|$ of 152.2 S/cm for the $\sigma_y$ component (see Fig.~\ref{fig:Piezospintronic}). On the other hand, the phase transition from the $\Gamma_{4g}$ to $\Gamma_{5g}$ states for Cr$_3$SnN leads to a $|$$\Delta$AHC$|$ as large as 127.6 S/cm (Fig.~\ref{fig:Piezospintronic}). This is comparable to the $|$$\Delta$AHC$|$ of 241.8 S/cm obtained assuming the $\Gamma_{4g}$ state with 1\% tensile strain. Therefore, we suspect that those compounds with $\Gamma_{4g}$ states are more promising to host a strong piezospintronic effect.
\subsection{Anomalous Nernst effect}
The intriguing behavior of AHC under biaxial strain can be attributed to the fine-tuning of the electronic structure. To further illustrate such sensitivity, we investigated the ANC, which shares the same symmetry as AHC but is proportional to the derivative of AHC with respect to energy following the Mott formula.~\cite{xiao2006berry} Taking Mn$_3$NiN as an example, the corresponding ANC is as large as 16569 S/cm.eV. This is consistent with 1.80 AK$^{-1}$m$^{-1}$ at 200K obtained by Zhou {\it et al.}~\cite{zhou2020giant}, up to a factor of two but with alike dependence with respect to the energy around the Fermi energy (see Fig. S15). The numerical discrepancy might be due to the different lattice constants used in the calculations, as discussed above. Such a large ANC of Mn$_3$NiN is comparable to the largest values observed experimentally in Co$_2$FeGe (3.16 AK$^{-1}$m$^{-1}$), Co$_2$MnGa (4.0  AK$^{-1}$m$^{-1}$), and Fe$_3$Ga (3.0  AK$^{-1}$m$^{-1}$).~\cite{sakai2018giant, guin2019anomalous, noky2018characterization, sakai2020iron} Interestingly, the strain also has a stronger influence on the ANC. For the tetragonally distorted Mn$_3$NiN in the $\Gamma_{4g}$ state, the ANC becomes as large as 20035 S/cm.eV at -0.5\% compressive strain for the $\alpha_x$ and $\alpha_y$ components, {\it i.e.}, enhanced by 21\% compared to that of the cubic case. Moreover, the ANC of Cr$_3$SnN is enhanced significantly, resulting in a value of -16321 S/cm.eV at 0.5\% tensile strain for the $\alpha_z$ component, which is comparable to that in Mn$_3$NiN. Last but not least, a sign change can be induced in the ANC by the biaxial strain. For instance, the ANC corresponding to the $\alpha_x$ and $\alpha_y$ components of Cr$_3$XN changes to -3196.9 S/cm.eV at 1\% compressive strain starting from 6823.4 S/cm.eV for the cubic case in the $\Gamma_{4g}$ state. This is also observed for other APVs such as Cr$_3$XN (X=Ir and Sn) and Mn$_3$XN (X=Co, Rh, Ag, Sn, and Ir) (see Fig. S11).
\\
To explicate the origin of the piezospintronic effects on both AHC and ANC, our detailed analysis of the electronic structure reveals that the tunability of AHC and ANC by strain can be attributed to the presence of Weyl points close to the Fermi energy. Taking Mn$_3$PdN as an example, as shown in Fig.~\ref{fig:ANE}(a), the ANC can be tuned between 7037.3 S/cm.eV at 1\% compressive strain to -14599 S/cm.eV at 1.5\% compressive strain, with a sign change at 1.3\% strain. Such an ANC is comparable to that observed in Mn$_3$NiN and Cr$_3$SnN, awaiting further experimental validation. The band structure shows no obvious changes due to the applied strain (see Fig.~\ref{fig:ANE}(b)). However, it is found that there are 12 Weyl points at (0.43, -0.35, -0.49) and equivalent k-points with an energy of 20 meV below the Fermi energy in the cubic phase. Such Weyl points will be shifted across the Fermi energy upon applying compressive strain, {\it e.g.}, reaches to -1.75 meV for -1.3\% biaxial strain, and 1.87 meV above the Fermi energy for 1.4\% strain (see above inset of Fig.~\ref{fig:ANE}(a)). The AHC is mostly enhanced when the Weyl points are located at the Fermi energy, due to the singular behavior of the Berry curvature at the Weyl nodes (see below inset of Fig.~\ref{fig:ANE}(a)). This is consistent with Ref. 51 and our observation in Mn$_3$GaN.~\cite{huyen2019topology, Samathrakis-Mn3GaN} For magnetic materials, the Weyl nodes with opposite chiralities ({\it i.e.}, Berry curvatures with opposite sign) are located at the same energy, but the total contributions to the AHC can be significant, particularly if the Weyl nodes are within ``several'' or ``a few'' meV around the Fermi energy. Correspondingly, the sign of ANC, which is determined by the derivative of AHC can be tuned as the Weyl points are shifted across the Fermi level. We suspect such tunability of AHC and concomitant ANC by manipulating the Weyl points with the biaxial strain or other possible stimuli is promising for future applications, such as Fe$_3$Al with giant transversal thermoelectric effects.~\cite{sakai2020iron}
\section{Conclusion}   
In summary, we carried out a systematic analysis of 16 cubic antiperovskite M$_3$XZ compounds with noncollinear magnetic ground states, focusing on the magnetic properties driven by isotropic and anisotropic magnetostructural coupling. It is found that there exists a strong competition between different noncollinear magnetic configurations where a large MAE clearly defines the noncollinear magnetic ground state, while small MAE leads to the mixed $\Gamma_{4g}$ and $\Gamma_{5g}$ configurations. For such materials with mixed magnetic ions, the MAE cannot be understood based on the perturbation theory, whereas the SOC energy can be used to get a reliable atomic resolved contribution, which is mostly determined by the strength of atomic SOC. The magnetic ground state analysis resulted in the prediction of five novel APVs with  noncollinear ground state, especially Cr$_3$XN APVs (X=Ir, Pt, and Sn) with the $\Gamma_{4g}$ magnetic configuration. The isotropic magnetostructural coupling indicated by the magnetic deformation $\Delta$a/a$_0$ can be considered as an effective descriptor for the magnetocaloric effect. However, we observed that the magnetic deformation is better measured comparing the paramagnetic and magnetically ordered states, resulting in better agreement with the experimental NTE results, rather than comparing the NM and ordered state. Therefore, we suggest the recently proposed proxy for predicting magnetocaloric effects based on the magnetic deformation could be improved by using quasi-random approximation of the paramagnetic state. More interestingly, biaxial strain not only causes a significant PME in such materials but also leads to a strong influence on MAE, AHC, and ANC after considering spin-orbit coupling. Based on detailed symmetry analysis, we performed an explicit evaluation of the AHC and found that those compounds with the $\Gamma_{4g}$ magnetic ground state have large AHC, which is susceptible to the epitaxial strain. Nevertheless, there is no strong correlation between the net magnetization and induced AHC when finite strain is applied. Detailed analysis reveals that the sensitivity of AHC and derived ANC with respect to strain can be attributed to the fine-tuning of energies for the Weyl nodes, opening up further possibilities for engineering spintronic devices in the future.
\section{Acknowlegments}
The authors are grateful and acknowledge TU Darmstadt Lichtenberg high-performance computer support for the computational resources where the calculations were conducted for this project. The authors thank Prof. Manuel Richter of IFW Dresden for providing the SOC strength data and discussion. This project was supported by the Deutsche Forschungsgemeinschaft (DFG, German Research Foundation)-Project-ID 405553726-TRR 270. Nuno Fortunato acknowledges European Research Council (ERC) funding for financial support under the European Union’s Horizon 2020 research and innovation programme (Grant No. 743116 project Cool Innov). The work of Jan Zemen was supported by the Ministry of Education, Youth and Sports of the Czech Republic from the OP RDE program under the project International Mobility of Researchers MSCA-IF at CTU No.CZ.02.2.69/0.0/0.0/18$_-$070/0010457. 
% \bibliography{my_file}
%\bibliography{achemso-demo}
%\bibliographystyle{ieeetr}

\end{document}